\documentclass{elsart}

\usepackage{graphics}
\usepackage{graphicx}
\usepackage{epsfig}

\usepackage{amssymb}
\def\lsim{\raise0.3ex\hbox{$<$\kern-0.75em\raise-1.1ex\hbox{$\sim$}}}
\def\gsim{\raise0.3ex\hbox{$>$\kern-0.75em\raise-1.1ex\hbox{$\sim$}}}

\newcommand{\be}{\begin{equation}}
\newcommand{\ee}{\end{equation}}

\def\beq{\begin{equation}}
\def\eeq{\end{equation}}
\def\beqa{\begin{eqnarray}}
\def\eeqa{\end{eqnarray}}

\def\gappeq{\mathrel{\rlap {\raise.5ex\hbox{$>$}}
{\lower.5ex\hbox{$\sim$}}}}

\def\lappeq{\mathrel{\rlap{\raise.5ex\hbox{$<$}}
{\lower.5ex\hbox{$\sim$}}}}

\def\Toprel#1\over#2{\mathrel{\mathop{#2}\limits^{#1}}}

\newcommand{\rk}{\mbox{\boldmath $k$}}

\begin{document}

\begin{frontmatter}

\title{Hadron and Photon Production in the Forward Region at RHIC and LHC}

\author[Marcos]{M. A. Betemps} and
\ead{marcos.betemps@ufpel.edu.br}
\author[Victor]{V. P. Gon\c calves}
\ead{barros@ufpel.edu.br}
\address[Marcos]{Conjunto Agrot\'ecnico Visconde da Gra\c ca, CAVG,\\
Universidade Federal de Pelotas,
Caixa Postal 460, CEP 96060-290, Pelotas, RS, Brazil}
\address[Victor]{Instituto de F\'{\i}sica e Matem\'atica,\\
Universidade Federal de Pelotas,
Caixa Postal 354, CEP 96010-900, Pelotas, RS, Brazil}
\begin{abstract}
In this paper we investigate the hadron and photon production at forward rapidities in $pp$ and $pA$ collisions considering two phenomenological models based on the  Color Glass Condensate formalism. The predictions of these models for the hadron spectra are
compared with the dAu and pp RHIC data and the nuclear modification factor $R_{hA}$ is estimated. Our results demonstrate that the analysis of the transverse momentum dependence of this factor allows to discriminate between the phenomenological models. Furthermore, we predict  $R_{hA}$ for photon production at forward rapidities at RHIC and LHC energies and study the $p_T$ dependence of the photon to pion production ratio.
\end{abstract}

\end{frontmatter}

\section{Introduction}
\label{intro}

Quantum Chromodynamics (QCD) at high energies can be described by an effective theory denoted Color Glass Condensate (CGC), which is a many-body theory of partons which are weakly coupled albeit non-perturbative due to the large number of partons (For reviews see Ref. \cite{hdqcd}). 
Properties of the CGC are specified by correlation functions of gluons which evolve with increasing energy. They obey an infinite hierarchy of non-linear evolution equations - the Balitsky-JIMWLK hierarchy \cite{BAL,CGC}. In the mean field approximation, the first 
equation in the  hierarchy decouples and it boils down to a single nonlinear integro-differential  equation: the Balitsky-Kovchegov (BK) equation \cite{BAL,KOVCHEGOV}. In particular, the BK equation determines in the large-$N_c$ limit the evolution of the two-point correlation function, which corresponds to  scattering amplitude ${\cal{N}}(Y,r)$ of a dipole off the CGC, with $r$ the dipole size and $Y \propto \ln s $ is the rapidity. This quantity  encodes information about the hadronic scattering  and thus about the non-linear and quantum effects in the hadron wave function. Although the general solution to the BK equation still is not known, approximate solutions have been constructed which separately cover the non-linear regime deeply at saturation and the linear regime, where $\cal{N}$  obeys the BFKL or DGLAP equation  \cite{dipolos,iim,kkt,dhj,Goncalves:2006yt,buw}. 
The transition among these regimes is 
specified  by a typical scale, which is energy dependent and is
called saturation scale $Q_{s}$ [$Q_{\mathrm{s}}^2 \propto A^{\alpha} x^{-\lambda}$]. Basically, the  main properties of the solutions of the BK equation  are: (a) for the interaction of a small dipole ($r
\ll 1/Q_{s}$), ${\cal{N}}(Y=\ln 1/x,r) \approx r^2$, implying  that
this system is weakly interacting; (b) for a large dipole
($r \gg 1/Q_{s}$), the system is strongly absorbed and therefore 
${\cal{N}}(Y,r) \approx 1$.  This property is associated  to the
large density of saturated gluons in the hadron wave function.  
Moreover, the BK equation predicts the geometric scaling regime: at small values of $x$, instead of being a function of a priori the two variables $r$ and $x$, ${\cal{N}}(Y,r)$ is actually a function of a single variable $r^2Q_s^2(x)$ up to inverse dipole sizes significantly larger than the saturation scale. This scaling is obvious at $r \gg 1/Q_{s}$, but it is a non-trivial prediction for $r < 1/Q_{s}$. Furthermore, it breaks down for $r \ll 1/Q_{s}$ (leading-twist regime), which implies a limited extension for the geometric scaling window. The scaling behavior is predicted to hold approximately in the range  $r_{gs} \lesssim r \lesssim r_s$, where $r_{gs} \approx 1/Q_{gs}$  ($Q_{gs} = Q_s^2/\Lambda$)  and $r_s \approx 1/Q_s$. The so-called  extended scaling region is characterized by the geometric scaling momentum $Q_{gs}$, which grows faster than the saturation  scale with $x$ and defines the upper bound in transverse momentum of the geometric scaling region.

 The search of signatures for the parton saturation effects has been an
active subject of research in the last years.
In particular,  the geometric scaling window has been observed in inclusive and diffractive processes at HERA \cite{scaling,marquet,prl}   and the  observed \cite{brahms,star}
suppression of high $p_T$ hadron yields at forward rapidities in $dAu$ collisions at RHIC 
{ had} its behavior  anticipated on the basis of CGC ideas \cite{cronin}.
 A current shortcoming of these analyzes comes from the  non-existence of an exact solution of 
the non-linear equation in the full kinematic range, which implies the construction of 
phenomenological models satisfying the asymptotic behavior which { is} under theoretical 
control.   Several models for the forward dipole cross section have been used in the literature 
in order to fit the HERA and RHIC data \cite{dipolos,iim,kkt,dhj,Goncalves:2006yt,buw}. In general, the  adjoint dipole scattering amplitude is modeled in the coordinate space in terms of a simple Glauber-like formula as follows
\beq
{\cal{N}_A}(x,r) = 1 - \exp\left[ -\frac{1}{4} (r^2 Q_s^2)^{\gamma (x,r^2)} \right] \,\,,
\label{ngeral}
\eeq
where   $\gamma$ is the anomalous dimension of the target gluon distribution. The fundamental scattering amplitude ${\cal{N}_F}$ can also be parameterized as in (\ref{ngeral}), with the replacement $Q_s^2 \rightarrow Q_s^2\, C_F/C_A = 4/9\, Q_s^2$. Moreover, it is useful to assume that the impact parameter dependence of $\cal{N}$ can be factorized as 
${\cal{N}_{F,A}}(x,r,b) = {\cal{N}_{F,A}}(x,r) S(b)$.  The main difference among the distinct phenomenological models comes from the  predicted behavior for the anomalous dimension, which determines  the  transition from the non-linear to the extended geometric scaling regimes, as well as from the extended geometric scaling to the DGLAP regime. It is  the behavior of $\gamma$ that determines the fall off with increasing $p_T$ of the cross section. The current models in the literature consider the general form $\gamma = \gamma_s + \Delta \gamma$, where $\gamma_s$ is the anomalous dimension at the saturation scale and $\Delta \gamma$ mimics the onset of the geometric scaling region and DGLAP regime. One of the basic differences between these models is associated to the behavior predicted for $\Delta \gamma$. While the models proposed in Refs. \cite{iim,kkt,dhj} assume that $\Delta \gamma$ depends on terms which violate the geometric scaling, i.e. depends separately on $r$ and rapidity $Y$,  the model recently proposed in Ref. \cite{buw} consider that it is a function of $r Q_s$.  In particular, these authors  demonstrated that the RHIC data for hadron production in $dAu$ collisions for all rapidities are compatible with geometric scaling and that 
 geometric scaling violations  are not observed at RHIC \cite{buw}. In contrast, the IIM analysis  implies that a substantial amount of geometric scaling violations is needed in order to accurately describe the $ep$ HERA experimental data \cite{iim}. These contrasting results motivate the study of other experimental observables directly associated to the CGC physics.

Our goal in this paper is twofold. Firstly, to compare the CGC predictions for the ratio $R_{hA}$ [see definition in Eq. (\ref{rha}) below] with the RHIC experimental data for hadron production.  Although the $p_T$ dependence of the hadron spectra in $dAu$ collisions is quite well described, the overall magnitude of the production cross section still is uncertain, which is associated to the fact that the CGC calculations have been performed at leading order. Moreover, the description of hadron production in $pp$ collisions at forward rapidities using   CGC physics  still  is an open question (See e.g. \cite{boer_spin}). These two aspects limit the predictive power of the CGC formalism for the behavior of the ratio $R_{hA}$. To overcome this, we extend the model proposed in Ref. \cite{buw} for $pp$ collisions and compare with the recent STAR data for inclusive $\pi^0$ production at $\eta = 3.3,\, 3.8$ and 4.0 \cite{star}. It allows to determine the magnitude of the next-to-leading order corrections in terms of the $K$-factor necessary to describe the $pp$ cross section. A similar calculation is performed for $dAu$ collisions, obtaining an identical value for the $K$-factor  at $\eta = 4.0$. It  allows  to predict the behavior and normalization of $R_{hA}$ for this rapidity.  Assuming that  this behavior also is present for other rapidities and  for charged hadron production, we  compare the CGC predictions with the BRAHMS data \cite{brahms}. Our results demonstrate that the study of the $p_T$ dependence of the ratio $R_{hA}$ allows to discriminate between the distinct phenomenological models.

Our second goal is  to present the predictions of the CGC physics for photon production using a model for the scattering dipole amplitude which describes quite well the hadron production. As emphasized in Refs. \cite{jamal_photon,betemps,jamal_tam}, it is essential to consider the electromagnetic probes of the CGC in order to determine the dominant physics in the forward region at RHIC and LHC. Distinctly from hadron production,  there is no hadronization of the final state present in the description of the photon production cross section, which implies that it is a cleaner probe of the CGC. We estimate the ratio $R_{hA}$ for photon production at forward rapidities for RHIC and LHC energies and compare its behavior with that predicted for hadrons. Moreover, as a by product, we estimate the 
photon to pion production ratio and study its $p_T$ dependence.

The paper is organized as follows. In the next section (Section \ref{hp}) we briefly review the hadron and photon production in the Color Glass Condensate formalism and the main characteristics of the distinct parameterizations  for the dipole amplitude scattering. In Section \ref{rpa_hadrons} we define the nuclear modification factor $R_{hA}$ and discuss the theoretical expectations for the behavior of this ratio. Moreover, we calculate  the inclusive $\pi^0$ production in $pp$ collisions using the CGC formalism and estimate the ratio $R_{hA}$ for pions and charged hadrons. Our predictions are compared with the STAR and BRAHMS data. The ratio $R_{hA}$ for photon production is estimated in Section \ref{rpa_photons} and the photon to pion production ratio is calculated.  Finally, in the Section \ref{conclu} our  main conclusions  are summarized.

\section{Hadron and Photon Production in the Color Glass Condensate Formalism}
\label{hp}

Lets consider the hadron production at forward rapidity in $dAu$ collisions. As pointed in Ref. \cite{difusivo}, it is  a typical example of a dilute-dense process, which is an ideal system to study the small-$x$ components of the $Au$ target wave function.  In this case the cross section is expressed as a convolution of the standard parton distributions for the dilute projectile, the dipole-hadron scattering amplitude (which includes the high-density effects) and the parton fragmentation functions.  Basically, the minimum bias invariant yield  for single-inclusive hadron production in hadron-hadron  processes is described in the CGC formalism  by \cite{dhj,arata}
\begin{eqnarray}\nonumber
 \frac{d^2N^{pp(A)\rightarrow hX}} {dyd^2p_T}&=& \frac{1}{(2\pi)^2}
\int_{x_F}^1 dx_1 \frac{x_1}{x_F}\left[ f_{q/p}(x_1,p_T^2){\cal{N}_F}\left(x_2,\frac{x_1}{x_F}p_T\right)
D_{h/q}\left(\frac{x_1}{x_F},p_T^2\right)
\right.\\
&+& \left.   f_{g/p}(x_1,p_T^2){\cal{N}_A}\left(x_2,\frac{x_1}{x_F}p_T\right)D_{h/g}\left(\frac{x_1}{x_F},p_T^2\right)  \right].
\label{eq:final}
\end{eqnarray}
where $p_T$, $y$ and $x_F$ are the transverse momentum, rapidity and the Feynman-$x$
of the produced hadron, respectively. The variable $x_1$ denotes the momentum
fraction of a projectile parton,    $f(x_1,p_T^2)$  the projectile parton
distribution functions  and $D(z, p_T^2)$ the parton fragmentation
functions into hadrons. These quantities  evolve according to the 
DGLAP~\cite{dglap} evolution equations and respect the momentum
sum-rule. In Eq. (\ref{eq:final}), ${\cal{N}_F}(x,\rk)$  and  ${\cal{N}_A} (x,\rk)$ 
are the fundamental and adjoint representations of the  forward dipole amplitude in  
momentum space, which represent the probability for scattering of a quark and a gluon off the nucleus, respectively.  Moreover, $x_F=\frac{p_T}{\sqrt{s}}e^{y}$ and the momentum fraction of the target partons is given by $x_2=x_1e^{-2y}$ (For details see e.g. \cite{arata}).

The photon production can be evaluated in a similar way \cite{jamal_gelis,jamal_photon}, and the minimum bias invariant yield  can be written in the form \cite{jamal_photon}
\begin{eqnarray}
 \frac{d^2N^{pA\rightarrow \gamma X}} {dyd^2p_T}\!=\!\frac{1}{(2\pi)^2} 
\int_{x_F}^1 \!dx_1\! \frac{x_1}{x_F}\!\left[f_{q/p}(x_1,p_T^2){\cal{N}_F}\left(x_2,\frac{x_1}{x_F}p_T\right)
\!D_{\gamma/q}\left(\frac{x_1}{x_F},p_T^2\right)
\right]\,\,, \nonumber \\
\label{eq:fotons}
\end{eqnarray}
where $p_T$ and $y$ are now the transverse momentum and rapidity of the produced photons. In this equation, $D_{\gamma/q}$ is the quark-photon fragmentation function. Distinctly from hadron production, the rate of photon production only depends of the quark content on the projectile hadron. It implies that for the region where the gluon contribution can be disregarded in hadron production, the behavior for hadron and production is expected to be similar \cite{jamal_photon}. Furthermore, the two cross sections are dependent on the fundamental dipole scattering amplitude, which is a building block of the CGC formalism.  Therefore, if ${\cal{N}_F}$ is constrained for instance  in hadron production, the calculation of the photon production is straightforward.

The Eqs. (\ref{eq:final}) and (\ref{eq:fotons}) are only applicable to forward/backward  rapidities in $pp$ collisions. On the other hand, in hadron-nucleus collisions at high energies, due to the $A$ dependence of the saturation scale, they are expected to also be valid for mid-rapidity.
It is important to emphasize that the minimum bias cross sections  discussed in our paper are obtained by impact-parameter averaging the inclusive hadron/photon production cross section, which in the CGC formalism depends on the impact parameter only through the saturation scale. In  Ref. \cite{arata} the author discuss two alternatives  to implement this calculation,  which implies  different values for the effective  saturation scale  $\langle Q_s^2 \rangle$ in minimum bias collisions. In what follows we assume that $\langle Q_s^2 \rangle = A_{eff}^{1/3} Q_0^2 (x_0/x_2)^{\lambda}$, with $A_{eff} = 18.5 \, (20.0)$ for $dAu$ ($pPb$) collisions, in order to compare our predictions with those obtained in Refs. \cite{dhj,buw}. Moreover, as in Ref. \cite{boer_spin}, we assume $A_{eff} = 1$ in the $pp$ case.

The basic input for the calculations of the hadron and photon production are the dipole scattering amplitudes ${\cal{N}_A}$ and ${\cal{N}_F}$, which are solutions of the Balitsky-JIMWLK hierarchy or the BK evolution equation in mean-field approximation.  As already explained in the introduction, the
general solution to the BK equation still is not known, which implies that is necessary to consider 
phenomenological models, based on CGC physics, in order to calculate the observables. In what follows we consider two distinct phenomenological models constructed to describe the RHIC data: the DHJ model   \cite{dhj} and the recently proposed BUW model  \cite{buw}. In these two models the adjoint dipole scattering amplitude in the momentum space is given by
\begin{eqnarray}
{\cal{N}_A}(x,p_T)= - \int d^2 r e^{i\vec{p_T}\cdot \vec{r}}\left[1-\exp\left(-\frac{1}{4}(r^2Q_s^2(x))^{\gamma(p_T,x)}\right)\right]\,\,,
\end{eqnarray}
where $\gamma$ is assumed  a function of $p_T$ rather than $r$ in order to compute the Fourier transform more easily. The fundamental scattering amplitude ${\cal{N}_F}$ is obtained from ${\cal{N}_A}$ by the replacement $Q_s^2 \rightarrow Q_s^2\, C_F/C_A = 4/9\, Q_s^2$. 
Moreover, in these models it is assumed that $\gamma(p_T,x)=\gamma_s+\Delta\gamma(p_T,x)$, with  $\gamma_s = 0.628$. In  the DHJ model,  $\Delta\gamma(p_T,x)$ is given by \cite{dhj}
\begin{equation}
\Delta \gamma_{DHJ}(p_T,x)=(1-\gamma_s)\frac{\log(p_T^2/Q_s^2(x))}
{\lambda y+d\sqrt{y}+\log(p_T^2/Q_s^2(x))},
\label{gam_dhj}
 \end{equation}
with $y = \log (1/x)$, $\lambda = 0.3$ and $d = 1.2$. On the other hand, in the BUW model \cite{buw}
\begin{eqnarray}
\label{BUWeq}
\Delta \gamma_{BUW} (p_T,x)=(1-\gamma_s)\frac{(\omega^a-1)}{(\omega^a-1)+b},
\end{eqnarray}
where $\omega  \equiv p_T/Q_s(x)$ and the two free parameters $a=2.82$ and $b=168$ are fitted in order do describe the RHIC data on hadron production. The main difference between the parameterizations is the presence of terms in the DHJ model which violate the  geometric scaling. Distinctly from the BUW model, which assumes that $\Delta \gamma$ satisfies the geometric scaling property,  the DHJ one predicts that it  behaves as $\log(p_T^2/Q_s^2(x))/y$ for large $y$ and $p_T^2 > Q_s^2$, violating the geometric scaling. Another important difference is that the large $p_T$ limit of $\gamma \rightarrow 1$ is approached much faster in the BUW model than in the DHJ one, which implies different predictions for the large $p_T$ slope of the hadron and photon yield. As shown in \cite{buw}, both models describe quite well the  $dAu$ RHIC data for forward rapidities ($y \ge 2.2$), but the DHJ model fails to describe the large $p_T$ data for smaller rapidities, where the $x_2$ values probed are not very small. In  next section we extend this analysis for $pp$ collisions and calculate the nuclear modification factor $R_{hA}$ for hadron production.

\section{The Nuclear Modification Factor for Hadron Production}
\label{rpa_hadrons}

In order to disentangle the nuclear medium effects it is useful to compare the data from hadron-nucleus ($hA$) collisions to proton-proton ($pp$) using the nuclear modification factor $R_{hA}$ defined as:
\begin{eqnarray}
R_{hA} = \frac{1}{N_{coll}} \left( \frac{d^2N^{hA}}{dy  d^2p_T}/\frac{d^2N^{pp}}{dy  d^2p_T}\right) \,\,,
\label{rha}
\end{eqnarray}
where $N_{coll}$ is the number of binary collisions at a given centrality in a $hA$ collision, which is obtained from a Glauber model calculation (See e.g. Appendix I in Ref. \cite{arleo}). In the absence of nuclear effects, hard processes, as for example hadron and photon production at large transverse momentum, scale with the number of binary collisions, which implies that $R_{hA} = 1$ in this case \cite{arleo,accardi_hp}. The rapidity behavior of this ratio is directly related to the description of the target. If the target is a dilute system, $R_{hA}$  is expected to grow with rapidity because the parton associated to the identified hadron  has interacted with a greater number of gluons, each contributing with a finite amount of transverse momentum. In this case we expect that the ratio assumes a value greater than one beyond some value of $p_T$. On the other hand, if the target is a saturated system, the ratio is expected to decrease in value for larger rapidities, since the CGC physics implies a reduction of the nuclear gluon distribution associated to the non-linear evolution.

Lets  present a brief review of the main theoretical expectations for the behavior of $R_{hA}$ associated to CGC physics (For a detailed review see, e.g, \cite{iancu}). At central rapidities, it is predicted the presence of a Cronin peak, which is  interpreted as reflecting the classical saturation and is understood as a result of Glauber-like multiple scattering off the gluon distribution produced by uncorrelated valence quarks.  The Cronin peak is predicted to  disappears after a short evolution in rapidity and $R_{hA}$ is suppressed, stabilizing  at a small value which approaches  one asymptotically  at large $p_T$. The rapid suppression of the ratio with increasing rapidity has been interpreted as a consequence of the strong difference  between the quantum evolution of the nucleus and that of the proton.  Since for a fixed value of $y$ and $p_T$ the proton and nucleus saturation scales are different, the transverse phase space available  for the evolution is larger for the proton than for the nucleus. The amount of suppression is estimated as being $R_{hA} \approx 1/N_{coll}^{1 - \gamma}$, where $\gamma$ is the anomalous dimension which depends on the rapidity and transverse momentum. Therefore $\gamma$ determines the maximal possible suppression of the nuclear modification factor due to the saturation effects. As the phenomenological CGC-based models assume different behaviors for the anomalous dimension,  the analysis of $R_{hA}$ can be useful to constrain the QCD dynamics.

\begin{figure}[t]
\begin{center}
\scalebox{0.55}{\includegraphics{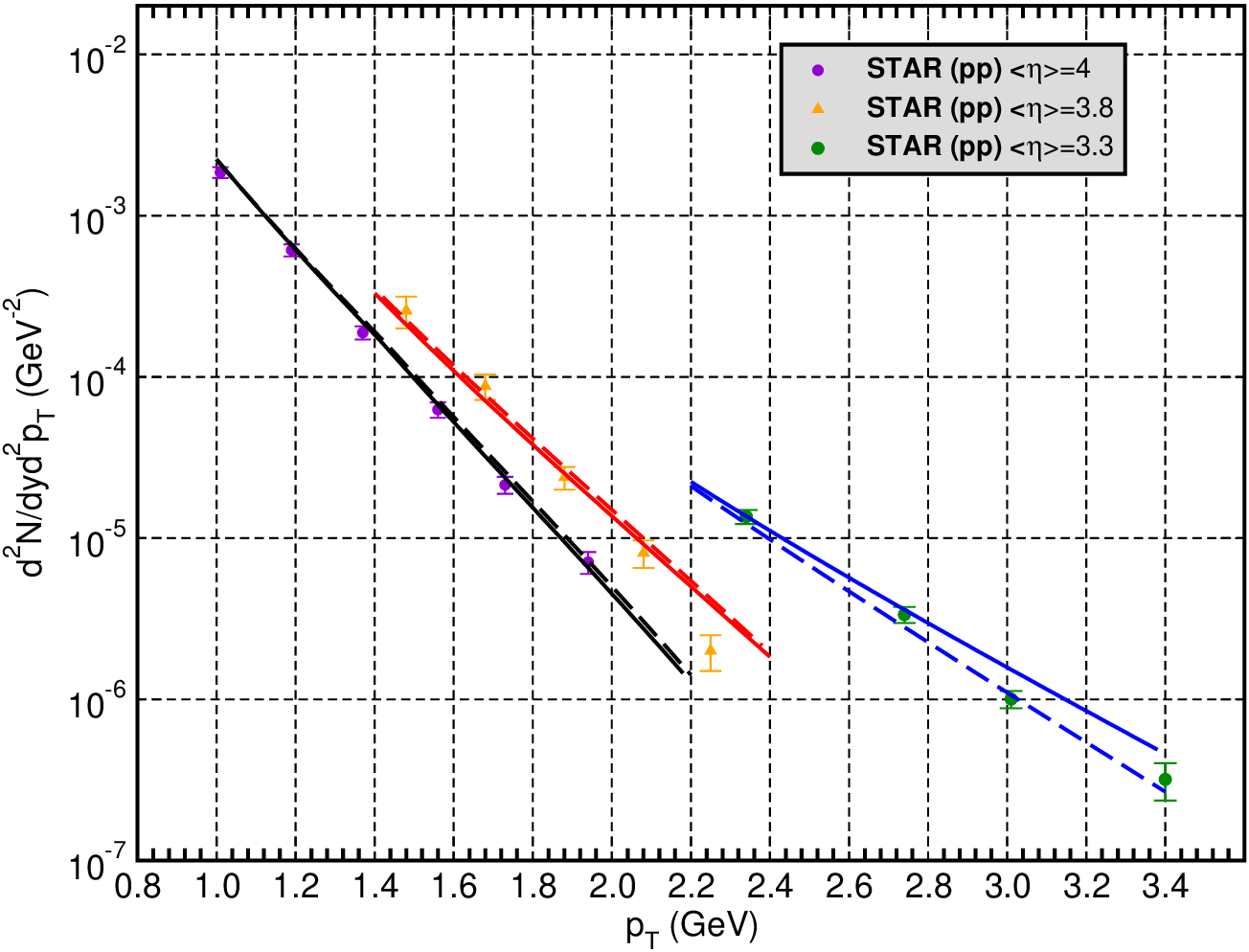}}
\end{center}
\caption{Inclusive $\pi^0$ production cross section in $pp$ collisions at RHIC energies. Data from STAR collaboration \cite{star}. We assume $K (\eta = 4.0) = K (\eta = 3.8) = 1.4$ and $K (\eta = 3.3) = 1.0$ for the DHJ (solid line)  and BUW (long-dashed line) predictions.} 
\label{fig1}
\end{figure}

The RHIC data for $R_{dAu}$ \cite{brahms} confirm the qualitative expectations of CGC physics \cite{cronin}. Although it is a very important evidence for CGC physics, it is fundamental to demonstrate the quantitative agreement  of the experimental data with the CGC predictions. In Ref. \cite{kkt} the authors have obtained a satisfactory description of the BRAHMS data for $R_{dAu}$ 
 assuming that a CGC-based description of high-$p_T$ hadron production in $pp$ collisions is valid (See also \cite{wied_scaling}). This is a strong assumption which should be verified. In principle, it is expected that for large rapidities the proton saturation scale assumes a large value, which implies a large value for the geometric scaling momentum $Q_{gs}$. Therefore, in this range the extended geometric scaling window becomes large and eventually covers the entire regime of particle production, since the DGLAP region is cut-off by energy-momentum conservation constraints \cite{boer_spin}. On the other hand, for mid-rapidity a CGC-based description for $pp$ collisions may not be well-justified. Consequently, it is important to test the applicability of the   CGC physics in $pp$ collisions at RHIC and verify the rapidity range in which this approach can be used. Recently, the STAR collaboration \cite{star} has reported the measurements of the production of forward $\pi^0$ mesons in $pp$  and $dAu$ collisions at $\sqrt{s_{NN}} = 200$ GeV. These data are ideal to check the CGC predictions. In Fig. \ref{fig1} we compare the DHJ and BUW predictions for the minimum bias invariant yield with the STAR $pp$ data. In our calculations we use the CTEQ5L parameterization \cite{cteq}  for the parton distribution functions  and the KKP parameterization for the  fragmentation functions \cite{kkp}. As in previous calculations \cite{dhj,buw} there is  one free parameter in our calculation: the $K$-factor. It is determined in order to obtain the better description of the experimental data and is fixed for each rapidity. We can see that the DHJ and BUW predictions are almost identical at the two larger values of rapidity, but differ in the large $p_T$ region for $\eta = 3.3$, with the DHJ one being larger than the  data. This trend is similar to observed for $dAu$ collisions in \cite{buw} with increasing rapidity.

\begin{figure}[t]
\begin{center}
\scalebox{0.55}{\includegraphics{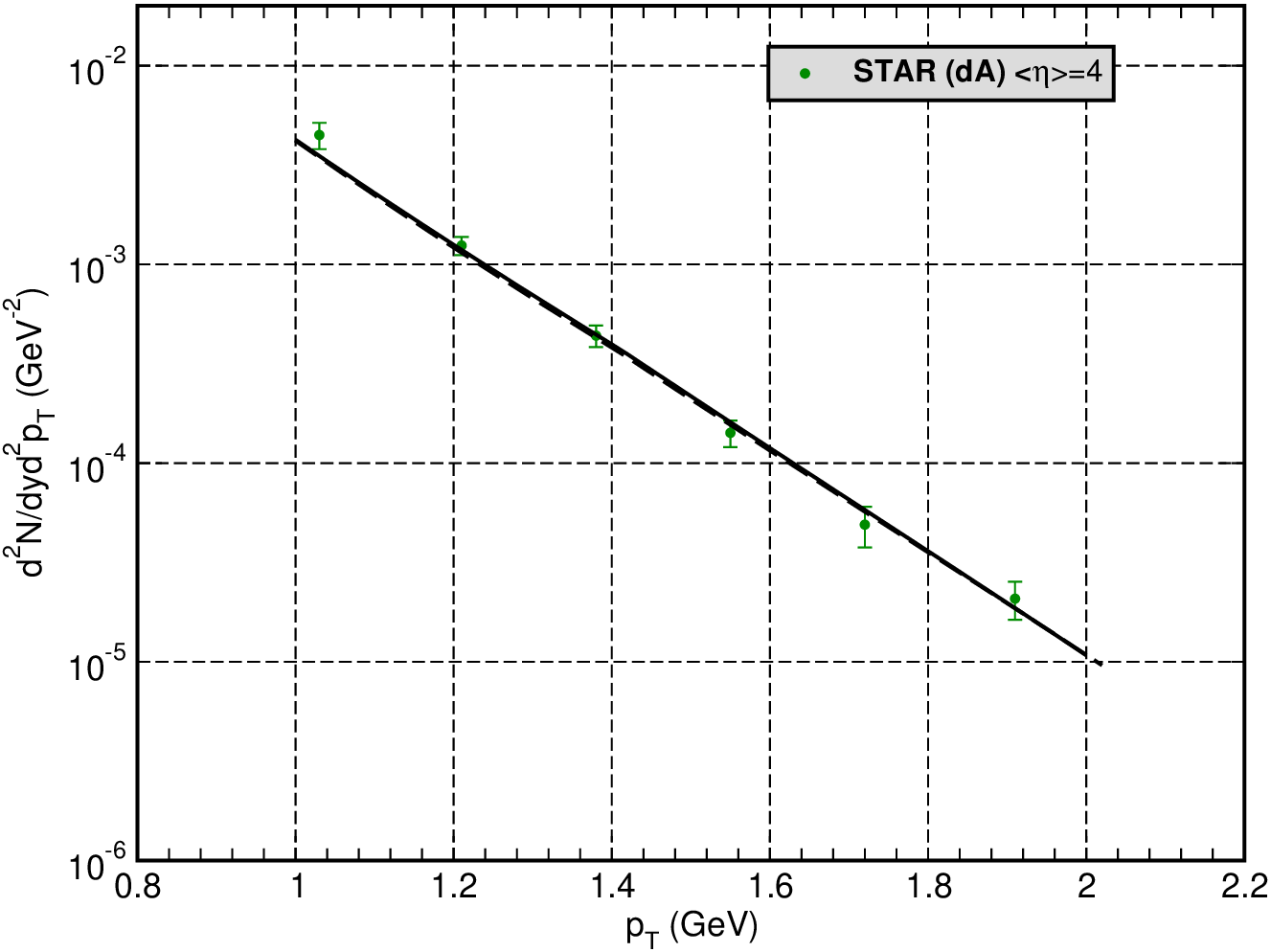}}
\end{center}
\caption{Inclusive $\pi^0$ production cross section in $dAu$ collisions at RHIC energies. Data from STAR collaboration \cite{star}. We assume $K (\eta = 4.0) = 1.4$ for the DHJ (solid line)  and BUW (long-dashed line) predictions.} 
\label{fig2}
\end{figure}

Some comments are in order here. First,  the $K$-factors for the different rapidities were fixed in order to describe the low-$p_T$ data, since they are in the extended scaling region where the formalism is expected to be valid (See Fig. 2 from Ref. \cite{boer_spin}). Second, the $K$-factor necessary to describe the experimental data at $\eta = 4$ is identical that found in Ref. \cite{boer_spin}, where the DHJ model was applied to describe $pp$ collision. Finally, we have found that $K(\eta = 3.3) < K(\eta = 4.0)$. This behavior is opposite to the observed when we apply the CGC formalism for charged hadron production in $dAu$ collisions. At this moment, we were not able to find a  reasonable explanation  for this particular behavior required to describe the  $pp$ data at $\eta = 3.3$.

\begin{figure}[t]
\label{fig3}
\begin{center}
\scalebox{0.55}{\includegraphics{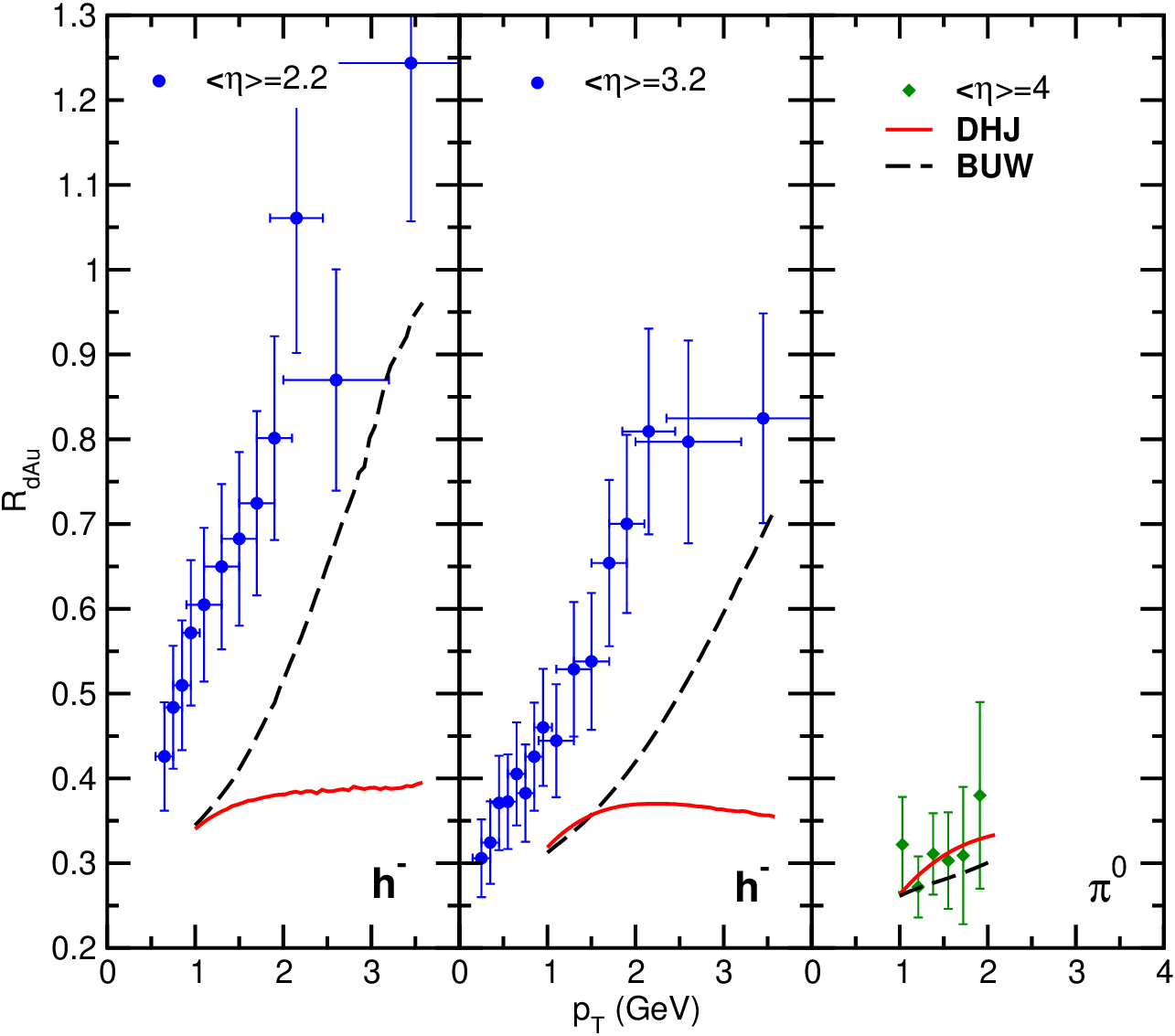}}
\end{center}
\caption{Nuclear modification ratio $R_{dAu}$ for charged hadrons and $\pi^0$ production at RHIC energies. Data from BRAHMS \cite{brahms} and STAR \cite{star} collaborations.}
\end{figure}

A similar study can be performed for the $\pi^0$ production in $dAu$ collisions. As in Refs. \cite{dhj,buw} we assume isospin invariance to obtain the parton distributions for a deuteron from those for a proton. In Fig. \ref{fig2} we present a comparison between the DHJ and BUW predictions and the STAR data for $\eta = 4$. We have that the two CGC-based predictions are very similar in this range, as already verified in \cite{buw}. An interesting aspect is that the $K$-factor necessary to describe the $dAu$ data is identical to that used in the description of the $pp$ data at the same rapidity. It implies that the resulting CGC prediction for the  ratio $R_{dAu}$ at this rapidity would be independent of the $K$-factor. Moreover, the $p_T$-behavior of this ratio would be a robust prediction of the CGC approaches. In Fig. \ref{fig3} (right panel) we present our predictions for the ratio $R_{dAu}$ for $\pi^0$ production and $\eta = 4$, where we have assumed that $N_{coll} = 7.2$ as useful in the experimental analysis \cite{brahms}. We have that the normalization and the $p_T$ dependence of the experimental data are quite well described by the CGC-based predictions. It is a strong evidence for the CGC physics in the forward rapidity at RHIC. However, in order to discriminate between the DHJ and BUW predictions we need to consider a larger range of rapidities.

 Motivated by the satisfactory description of the $\pi^0$ data in $pp$ collisions we extend our analysis for charged hadron production at  $\eta = 2.2$ and $3.2$, where we still expect that a CGC calculation is valid.  A current shortcoming is that there are not experimental data available in literature for charged hadron production at forward rapidities in $pp$ collisions. Therefore, it is not possible to constrain the $K$-factor for these cross sections. On the other hand, for $dAu$ collisions, the charged hadron spectra were studied by the BRAHMS collaboration \cite{brahms}. We have calculated the corresponding cross section and verified that the BUW model describe quite the data, while the DHJ model fails for central rapidity, as already verified in \cite{buw}. The basic difference between our results and those from \cite{buw} is that we have found a $K$-factor which is  two times larger than that obtained in \cite{buw}, which is directly associated to the treatment for the deuteron contribution to the cross section. In our case we have assumed that this contribution is normalized by the atomic number. It explains the difference by a factor two of our $K$-factor at $\eta = 4$ and that quoted in \cite{buw}. A comment is order here. We have estimated the contributions of 
${\cal{N}_F}(x,\rk)$  and  ${\cal{N}_A} (x,\rk)$ for the charged hadron cross section in $dAu$ collisions considering the BUW model and observed that, similarly to the DHJ one, the ${\cal{N}_F}(x,\rk)$ contribution determines the large $p_T$ behavior of the cross section for forward rapidities at RHIC energy, while ${\cal{N}_A} (x,\rk)$ is the relevant contribution at mid-rapidity.

In order  to calculate the ratio $R_{dAu}$ for charged hadron production and to compare with the BRAHMS data \cite{brahms} we assume that the $K$-factor is the same in our $dAu$ and $pp$ calculations. This assumption is not trivial: as the saturation scale of the nucleus and the proton are distinct, different dynamical effects are being probed for a fixed rapidity. Consequently, the normalization of our calculations of $R_{dAu}$ for charged hadron can be modified in the future. On the other hand, we believe that the  $p_T$ dependence predicted by the  CGC physics is a robust result which is directly associated to the anomalous dimension  considered in the distinct phenomenological models. In Fig. \ref{fig3} (left and middle panels) we present our predictions for  $R_{dAu}$ in charged hadron production using the DHJ and BUW models. We have that the  $p_T$ dependence predicted by these models is very distinct. While the DHJ model predict a ratio which is basically $p_T$ and $\eta$ independent, the BUW model predicts a strong $p_T$ dependence,  with $R_{dAu}$ increasing almost linearly with $p_T$, approaching one to large transverse momentum. Moreover, the BUW model also predicts a rapidity dependence for the ratio, with the slope increasing at smaller values of rapidity. These behaviors are observed in the experimental data. It is important to emphasize that both models describe the $dAu$ spectra for $\eta = 2.2$ and $3.2$ as shown in \cite{buw} and verified in our calculations. Consequently, the $p_T$ dependence of the ratio is directly associated to the distinct predictions for the $p_T$ spectra in $pp$ collisions, which are different already at $\eta = 3.3$, as verified in Fig. \ref{fig1}. The reasonable agreement between the BUW model and the experimental data is a strong evidence of the CGC physics. Moreover, it indicate that the dipole scattering amplitude satisfies the geometric scaling property in the forward RHIC kinematical range.

\begin{figure}[t]
\begin{center}
\scalebox{0.3}{\includegraphics{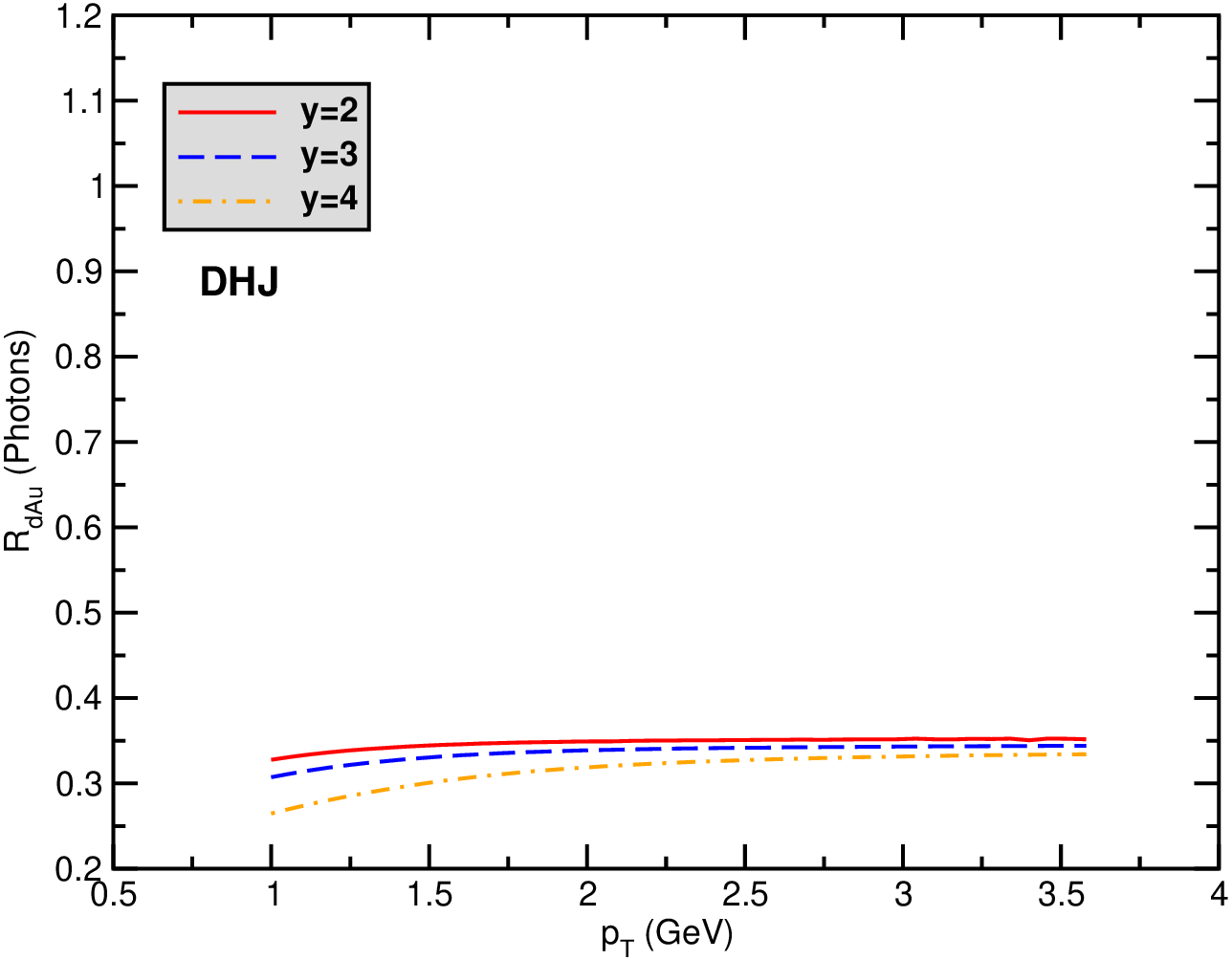}} \hspace{0.32cm}
\scalebox{0.3}{\includegraphics{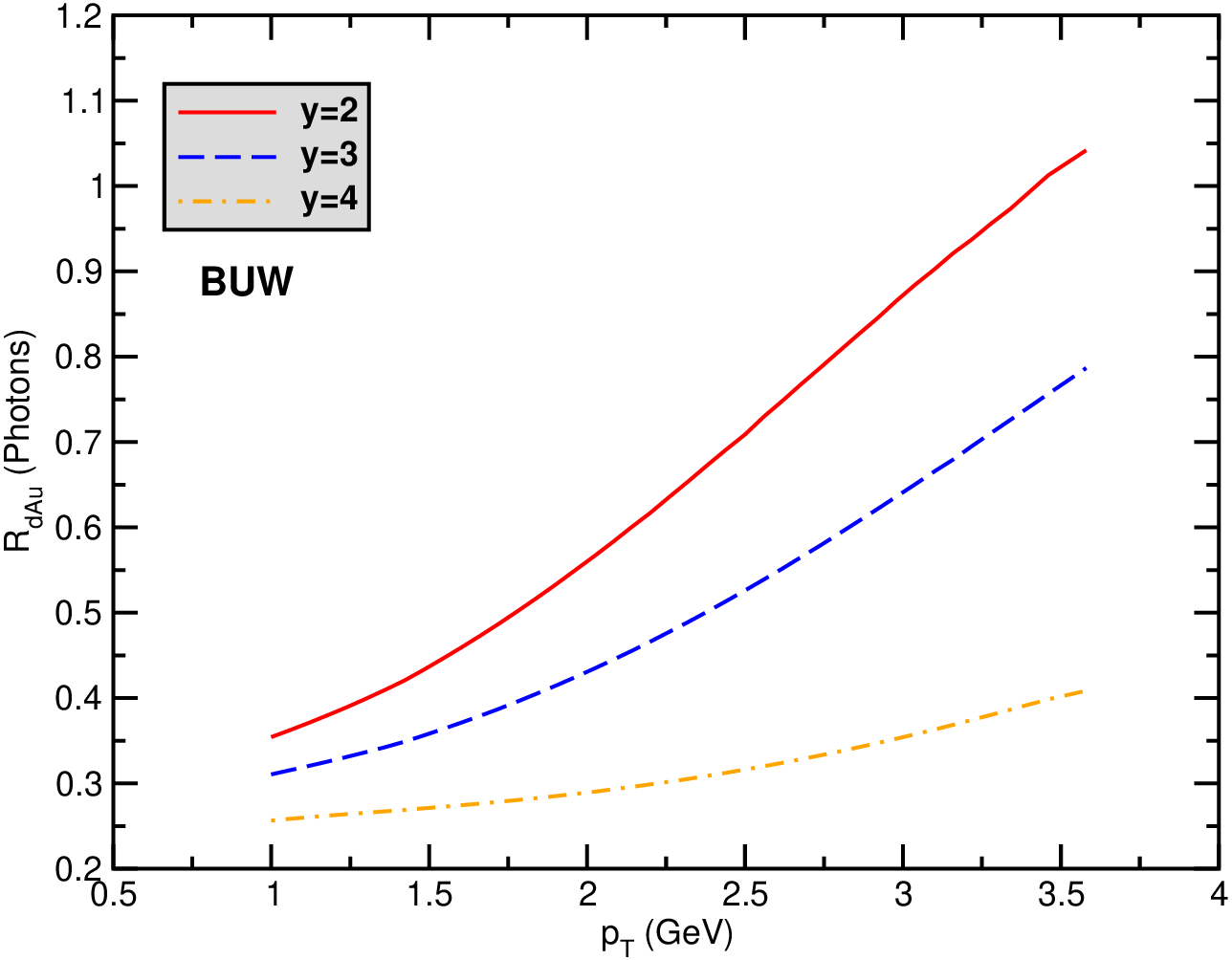}}
\end{center}
\caption{ Nuclear modification ratio $R_{dAu}$ for the photon production at RHIC energies.}
\label{fig4}
\end{figure}

\section{The Nuclear Modification Factor for Photon Production}
\label{rpa_photons}

The minimum bias yield for photon production in the CGC formalism can be calculated using the Eq. (\ref{eq:fotons}). The basic input is the fundamental scattering amplitude ${\cal{N}_F}$, which also is present in the calculations of hadron production cross sections. In particular, at forward rapidities it determines the behavior of this cross section, since the projectile  gluon distribution vanishes at $x_1 \rightarrow 1$. In the previous section we have estimated  the differential cross section for hadron production and obtained a  quite well description of the $p_T$ spectra for $pp$ and $dAu$ collision at forward rapidities, which implies that the behavior of ${\cal{N}_F}$ is reasonably well determined. It allows to obtain reliable predictions for  the behavior of the photon production cross section. Currently, 
experimental results at RHIC shown that the prompt photon cross section at mid-rapidity scale with $N_{coll}$ \cite{stankus}, which indicate that the nuclear effects are small at $\eta = 0$.
On the other hand, there is not  available experimental data for photon production at forward rapidities in $pp$ and $dAu$ collisions. 
We focus our analysis in the calculation of the ratio $R_{hA}$, as defined in the Eq. (\ref{rha}) above, at forward rapidities. Basically, we calculate the $pp$ and $hA$ minimum bias yields for photon production using Eq. (\ref{eq:fotons}), 
the CTEQ5L parameterization \cite{cteq}  for the parton distribution functions  and the GRV parameterization for the  quark-photon fragmentation function \cite{grv}. Similarly to hadron production we assume  that the $K$ factor is the same for $pp$ and $hA$ collisions. In the particular case of $dA$ collisions, we again assume $N_{coll} = 7.2$ and the isospin symmetry in order to calculate the parton distributions of deuteron (For a recent discussion about isospin effects in prompt photon production in $AA$ collisions see \cite{turbide}).

Initially lets calculate the ratio $R_{dAu}$ for photon production at RHIC energies ($\sqrt{s_{NN}} = 200$ GeV) and forward rapidities ($y = 2.0,\,3.0,\,4.0$). At smaller rapidities, a CGC description for $pp$ collisions is expected to breaks down. In Fig. \ref{fig4} we present our predictions for $R_{dAu}$ using the DHJ (left panel)  and the BUW model (right model). There is a large difference between the behaviors predicted by the two models. While the DHJ model predicts an almost flat ratio, which is $p_T$ and $y$ independent, the BUW model predicts that the ratio is flat only at very large rapidities, increasing with $p_T$ at smaller values of rapidity. This behavior is similar to that observed for hadron production. Consequently, the study of photon production can be an important search of information about the behavior of the scattering amplitude and the CGC physics.

\begin{figure}[t]
\begin{center}
\scalebox{0.55}{\includegraphics{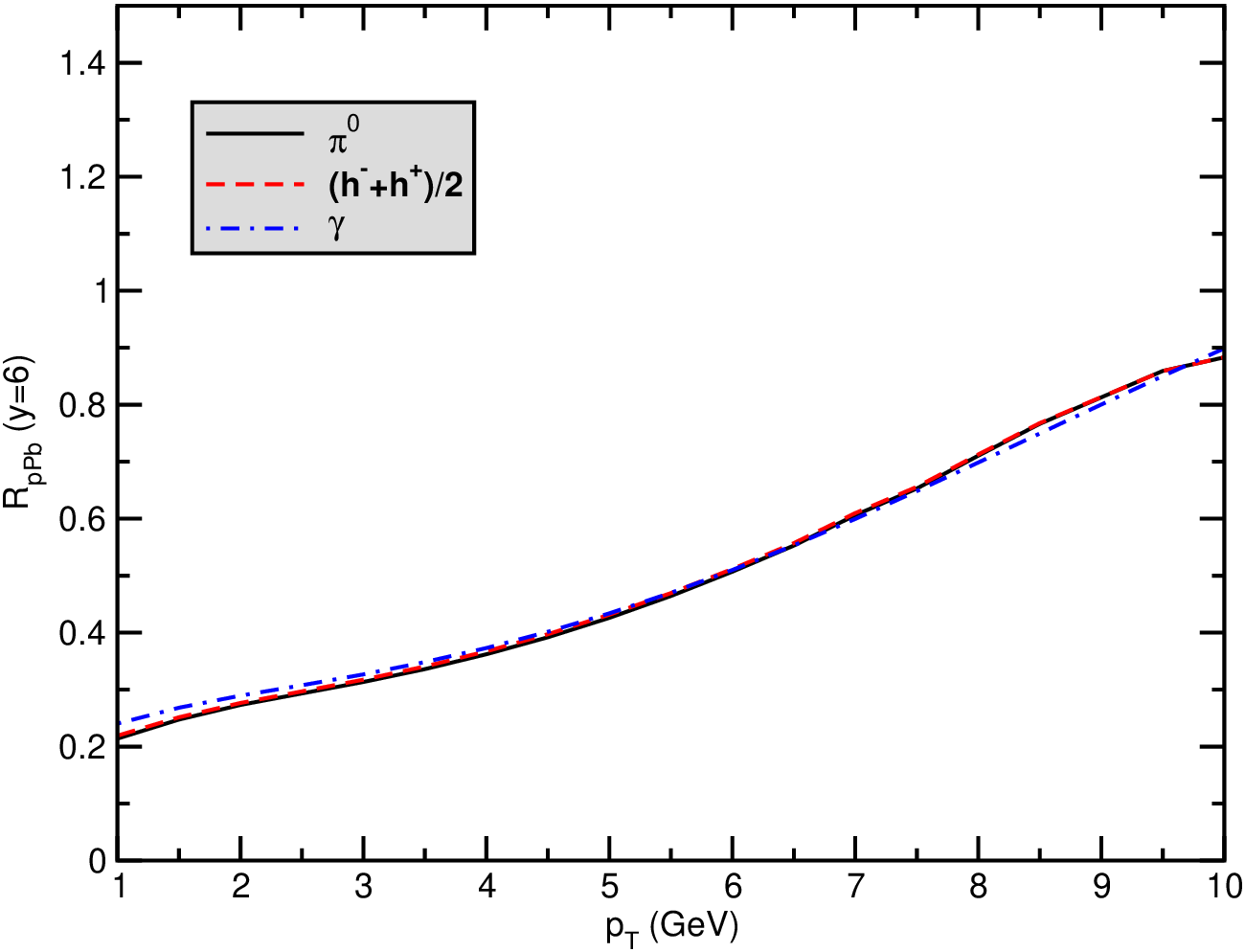}}
\end{center}
\caption{Nuclear modification ratio $R_{pPb}$ for charged hadrons, $\pi^0$ and photons at LHC energies ($\sqrt{s}=8.8$ TeV) and rapidity y=6, considering the BUW model.}
\label{fig_cinco}
\end{figure}

A shortcoming for the quantitative understanding of the CGC physics at RHIC is associated to the limited phase space in transverse momenta, which implies that the transitions expected to occur between the saturation, extended geometric scaling and  DGLAP regimes are not easily observed. In contrast, at LHC energies the available phase space will be much larger even at large rapidities, allowing to study the different regimes of the QCD at high energies in more detail.  Here we study the charged hadron, $\pi^0$ and photon production in $pp$ and $pPb$  collisions at $\sqrt{s}=8.8$ TeV and $y = 6$ using the BUW model and postpone a more detailed analysis for a future publication. For $pPb$ collisions we assume $A_{eff} = 20$ and  $N_{coll} = 7.4$ as quoted in the Table 6 of the  Appendix I from  \cite{arleo}.
In Fig. \ref{fig_cinco} we present our predictions for the ratio $R_{pPb}$. We can see that the magnitude and $p_T$ dependence is almost identical for the different observables. A similar result was obtained in Ref. \cite{tuchin}, where it was observed almost the same suppression as a function of transverse momentum for gluons and heavy quarks. Moreover,  we observe that the ratio increases with the transverse momentum, as already verified at RHIC. However, the ratio is almost one only at $p_T \ge 10$ GeV, which is directly associated to the larger window of the extended geometric scaling regime for the proton and nucleus at LHC energies.

\begin{figure}[t]
\begin{center}
\scalebox{0.3}{\includegraphics{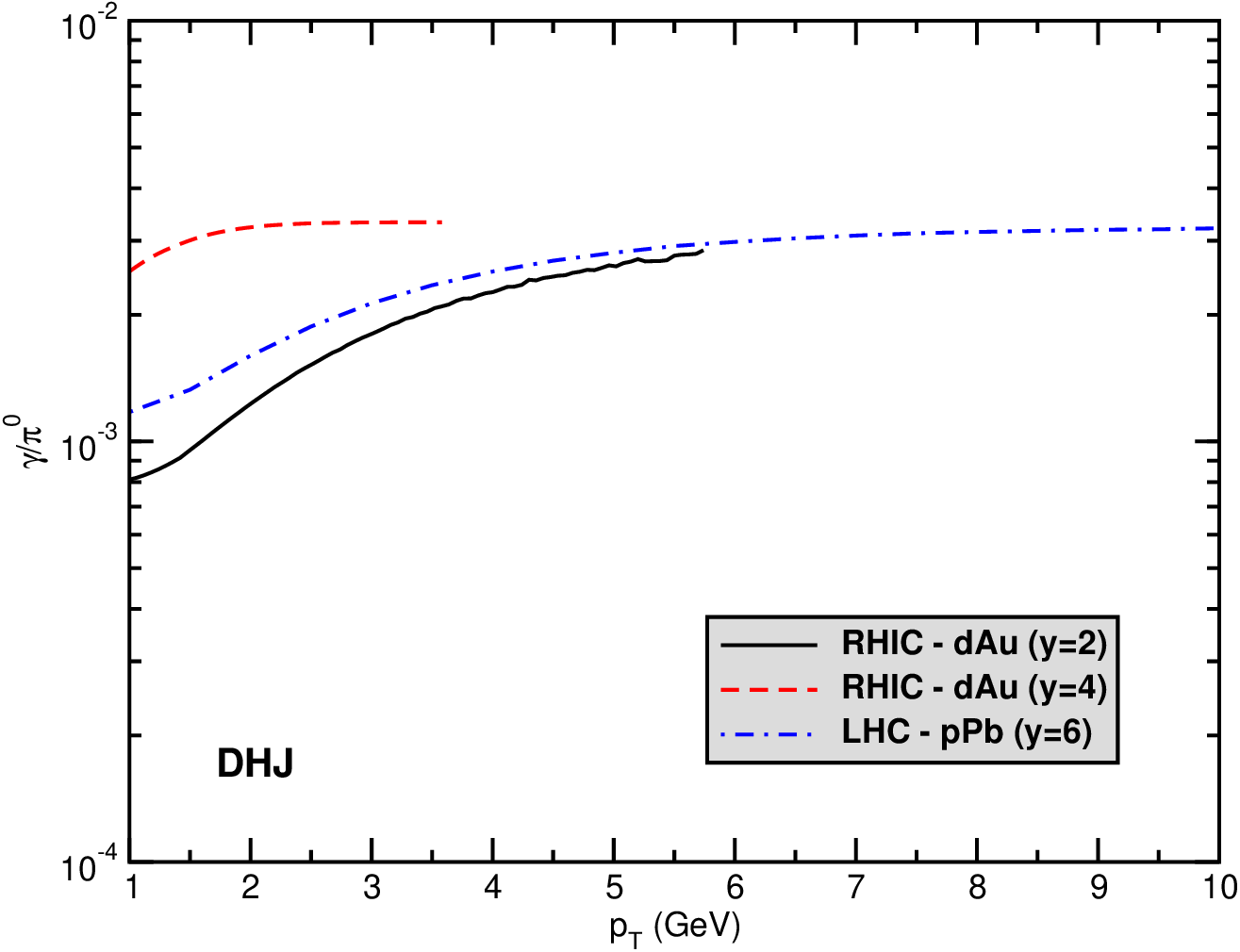}} \hspace{0.2cm}
\scalebox{0.3}{\includegraphics{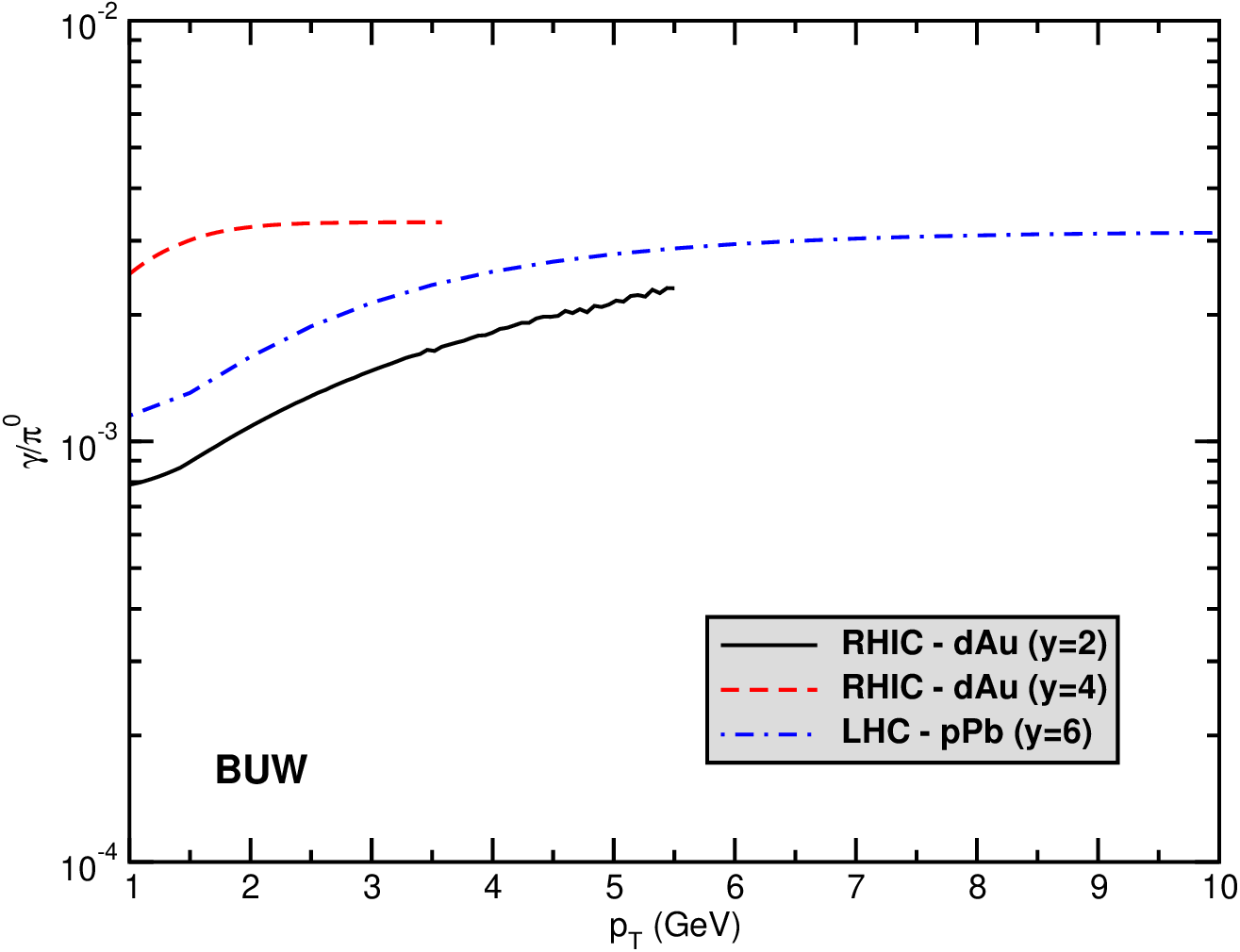}}
\caption{Ratio between the photon and $\pi^0$  production cross sections at RHIC energies, considering two values of rapidity at RHIC  ($y=2$ and $y=4$)  and $y=6$ at LHC energy.}
\end{center}
\label{last}
\end{figure}

Finally, as a by product we calculate the ratio between the photon and hadron cross sections. Distinctly from Ref.  \cite{jamal_ratio} we focus here in the transverse momentum dependence of this ratio at fixed rapidity for $dAu$ and $pPb$ collisions at RHIC and LHC energies. It is expected that the $K$-factor cancels in this ratio, which implies that its behavior should not be modified by next-to-leading corrections. Following  Ref. \cite{jamal_photon} we focus in the low $p_T$ region and forward rapidities, where the fragmentation contribution for photon production is expected to contributes significantly for the produced photons. In Fig. 6 we present our predictions for ratio $\gamma/\pi^0$, calculated using the Eqs. (\ref{eq:final}) and (\ref{eq:fotons}) considering the DHJ and BUW models. Due to the distinct phase space available for the different rapidities and energies, the curves in the figure finish in different points. We can see the DHJ and BUW results are similar, with  the ratio increasing in the small $p_T$ region and saturating at large values of the transverse momentum. It means that the ratio $\gamma/\pi^0$ is less sensitive  to the phenomenological model used as input in our calculations and is mainly determined by the photon and hadron fragmentation functions.  Moreover,  the ratio increases with the rapidity, as already verified in \cite{jamal_ratio}, which is associated to the fact that the gluon contribution in the projectile hadron diminishes with the rapidity. Finally, it is interesting to observe that the distinct predictions for the ratio tends to a same value at large $p_T$.

\section{Conclusions}
\label{conclu}
The observed suppression of the normalized hadron production  in $dAu$ collisions as compared to $pp$ collisions has been considered an important signature of the Color Glass Condensate physics. In the last years several models were proposed to describe the hadron spectra in $dAu$ collisions, obtaining a satisfactory description of these experimental data. In general, these models have been extended for $pp$ collisions  in order to calculate the ratio $R_{hA}$ without a comparison with the corresponding experimental data. In this paper we have,  for the first time, estimated the hadron production in $pp$ and $dAu$ collisions in a same theoretical formalism and compared these predictions with the experimental data. The comparison with the STAR data for $\pi^0$ production allows to fix the  free parameter in our calculations (the $K$-factor) and obtain a parameter free prediction for the nuclear modification ratio $R_{hA}$ at $\eta = 4$. 
For other rapidities, there are not available, simultaneously, $pp$ and $dAu$ experimental data for hadron production. In order to calculate the ratio $R_{hA}$ for these rapidities we have assumed that the $K$-factor is the same for $pp$ and $hA$ collisions. As discussed before, it is not a trivial assumption. However, the predictions for the $p_T$-dependence of the ratio are not affected by this choice.   
The comparison with the experimental data demonstrate that the BUW model, which assumes the geometric scaling property, is the adequate one for the RHIC kinematical range. 

We also have investigated the photon production at forward rapidities, which is considered a cleaner probe of the CGC physics. We demonstrate that the behavior of the ratio for photons is similar to verified for hadrons. It implies that the study of photon production is a useful search of information about the basic building block of the CGC formalism: the fundamental scattering amplitude. Finally, as a by product, we have estimated the ratio between the photon and hadron cross sections and demonstrated that is not sensitive to the phenomenological model used as input in the calculations.

\section*{Acknowledgments}
 This work was  partially financed by the Brazilian funding
agencies CNPq and FAPERGS.



\end{document}